\documentclass[pdflatex,sn-mathphys-num]{sn-jnl}

\usepackage{threeparttable}
\usepackage{graphicx}%
\usepackage{multirow}%
\usepackage{amsmath,amssymb,amsfonts}%
\usepackage{amsthm}%
\usepackage{mathrsfs}%
\usepackage[title]{appendix}%
\usepackage{xcolor}%
\usepackage{textcomp}%
\usepackage{manyfoot}%
\usepackage{booktabs}%
\usepackage{algorithm}%
\usepackage{algorithmicx}%
\usepackage{algpseudocode}%
\usepackage{listings}%
\usepackage{subcaption} 

\theoremstyle{thmstyleone}%
\theoremstyle{thmstyletwo}%

\theoremstyle{thmstylethree}%

\begin{document}

\title[Article Title]{Simulation of the electron escape ratio for keV alpha-particle ionization tracks in liquid helium}

\author*[1]{\fnm{Zihuai} \sur{Hu}}\email{cosmoses\_endeavour@163.com}

\author[1,2,3]{\fnm{Junhui} \sur{Liao}}\email{junhui.private@gmail.com, junhui\_private@163.com}
\equalcont{Communicating author.}

\author[1]{\fnm{Zhuo} \sur{Liang}}\email{liangzhuo\_w@163.com}

\author[1]{\fnm{Guangpeng} \sur{An}}\email{agp\_74@163.com}

\author[1]{\fnm{Zhaohua} \sur{Peng}} \email{pzh44@sina.com}

\author[1]{\fnm{Jian} \sur{Zheng}}\email{13522656935@139.com}

\author[4]{\fnm{Lifeng } \sur{Zhang}}\email{zlf20042008@126.com}

\author[4]{\fnm{Lei} \sur{Zhang}}\email{zlamp@163.com}

\author[5]{\fnm{Yuanning} \sur{Gao}}\email{yuanning.gao@pku.edu.cn}

\affil*[1]{\orgdiv{Department of Nuclear Physics}, \orgname{China Institute of Atomic Energy}, \orgaddress{\street{Sanqiang Rd. 1}, \city{Fangshan}, \postcode{102413}, \state{Beijng}, \country{China}}}

\affil[2]{\orgname{Yalong River Hydropower Development Company, Ltd}, \orgaddress{\street{288 Shuanglin Road}, \city{Chengdu}, \postcode{610051}, \state{Sichuan}, \country{China}}}

\affil[3]{\orgname{Jinping Deep Underground Frontier Science and Dark Matter Key Laboratory of Sichuan Province}, \orgaddress{\city{Liangshan}, \postcode{615000}, \state{Sichuan}, \country{China}}}

\affil[4]{\orgdiv{Department of Nuclear Synthesis Technology}, \orgname{China Institute of Atomic Energy}, \orgaddress{\street{Sanqiang Rd. 1}, \city{Fangshan}, \postcode{102413}, \state{Beijng}, \country{China}}}

\affil[5]{\orgdiv{School of Physics}, \orgname{Peking University}, \orgaddress{\street{ChengFu Rd. 209}, \city{Haidian}, \postcode{100084}, \state{Beijing}, \country{China}}}

\abstract{The electron escape ratio for 5.3 MeV alpha-particle ionization tracks in liquid helium under varying external electric fields has been measured in several experiments. However, to the best of our knowledge, the corresponding ratio for keV-scale alpha tracks in the same medium has not yet been reported. In this article, we demonstrate for the first time that this ratio can be accurately characterized using COMSOL-based simulations. Our simulation framework was developed in two stages. In Stage I, we aimed to verify consistency between our simulated results and published experimental data for 5.3 MeV alpha particles. Following successful verification in Stage I, we proceeded to Stage II, in which the 5.3 MeV track was replaced by 2, 5, and 10 keV tracks. Our simulated results reveal that (a) keV-scale tracks exhibit electron escape ratios approximately 1.5–2.5 times higher than that of the 5.3 MeV track, and (b) the escape ratios for all track energies (2, 5, 10 keV, and 5.3 MeV) exhibit a linear dependence on the ion number density at the simulation's T$_0$, but not on the electron number density.}

\keywords{WIMPs, Dark Matter Direct Detection, Liquid Helium,  Electron Escape Ratio, COMSOL}

\maketitle

\section{Introduction}
\subsection{Low-mass dark matter searches with ALETHEIA}
Although dark matter (DM) has been observed through compelling astrophysical evidence across all length scales, its particle nature remains elusive. Among the many proposed candidates, weakly interacting massive particles (WIMPs) are one of the most widely studied. Over the past decade or so, the experiments that implemented liquid xenon (LXe) time projection chambers (TPCs), such as XENON~\cite{XenonProject}, LUX/LZ~\cite{LZProject}, and PandaX~\cite{PandaXProject}, have played a leading role in the search for high-mass DM particles, roughly in the mass range of 10 GeV/c$^2$ to 10~TeV/c$^2$. In addition, liquid argon (LAr)-based TPCs, including those employed by the DarkSide~\cite{DarkSideProject} and DEAP~\cite{DeapProject} experiments, have also continued and will continue to push the boundaries on the upper limits of the DM-nucleon cross section in the same mass regime.

ALETHEIA (A Liquid hElium Time projection cHambEr In dArk matter) aims to hunt for low-mass dark matter ($\sim 100\text{s}~\mathrm{MeV/c}^{2}$ to $10\,\mathrm{GeV/c}^{2}$) with liquid helium-filled TPCs~\cite{ALETHEIA2023}. Similar to the successfully operating LXe and LAr TPCs in the community mentioned above, the current version of ALETHEIA TPCs will implement the ionization channel in the search of dark matter.

Over the past years, we have made significant progress and accumulated substantial experience that will help us in successfully building the dual-phase LHe TPCs. Our key achievements to date include the following: (i) cooling a house-made prototype LHe detector to approximately 4 K~\cite{ALETHEIA2023}; (ii) coating tetraphenyl butadiene (TPB)\footnote{LHe scintillation peaks at 80 nm (16 eV), which no commercial photosensor can detect directly. A wavelength shifter such as TPB is therefore necessary to convert this ultraviolet scintillation light into visible wavelengths suitable for detection.} on the inner walls of a cylindrical detector, with the TPB layer remaining functional even after several hours of cryogenic exposure at 4 K~\cite{TBPCoating4K}; (iii) proposing a novel ER and NR calibration method using helium beams~\cite{Gu2024}; and (iv) characterizing the key parameters of the NUV-HD-Cryo type SiPMs from Fondazione Bruno Kessler (FBK), including after-pulses, crosstalk, dark count rate, I–V curves, and photon detection efficiency, thereby confirming that these SiPMs are suitable as photosensors for LHe TPCs~\cite{SiPMs4KTests}.

\subsection{Electron escape ratio in LHe TPCs}

For a liquid noble gas dual-phase TPC, one of the critical parameters is the electron escape ratio, which represents the fraction of electrons that escape recombination with ions with the ``help'' of an external electric field, namely, the TPC's drift field. Once successfully escaped, the electrons are drifted toward the TPC's anode. In this article, we focus solely on the escape process itself.

A charged particle can ionize liquid helium to generate electron–ion pairs. For neutral particles such as neutrons and gammas, they first interact with the helium atom and subsequently produce alphas or electrons. The track shapes generated by these two types of particles are distinctly different. A 5.3 MeV alpha track exhibits a cylindrical shape, with a length of approximately 0.27 mm and a radius of approximately 60 nm~\cite{AstarWebsite, ito_effect_2012}. In contrast, the electron–ion pairs generated by a 365 keV electron have an average range of around 7 mm~\cite{EstarWebsite, Phan2020}. The mean distance between ionized pairs is approximately 2 nm for alpha tracks and approximately 840 nm for electron tracks. In this manuscript, we restrict our discussion to alpha tracks.

Although the escape ratio for 5.3 MeV alpha tracks has already been measured~\cite{gerritsen1948, williams1957}, to the best of our knowledge, no experimental or simulated ratio for keV-scale tracks, which correspond to ALETHEIA's region of interest (ROI), has yet been reported. It is well known that liquid helium undergoes a phase transition to a superfluid state once cooled to 2.17 K and below; however, the escape ratio exhibits a temperature-independent behavior between 1.4 K and 4.2 K under a 10 kV/cm field, according to reference~\cite{williams1957}. In our simulations, we set the temperature to 1.41 K to match the measured drift velocity under 10 kV/cm, as will be discussed further in section~\ref{sec:implementSimulation}. 

\begin{figure}[h]
	\centering
	\includegraphics[width=0.9\textwidth]{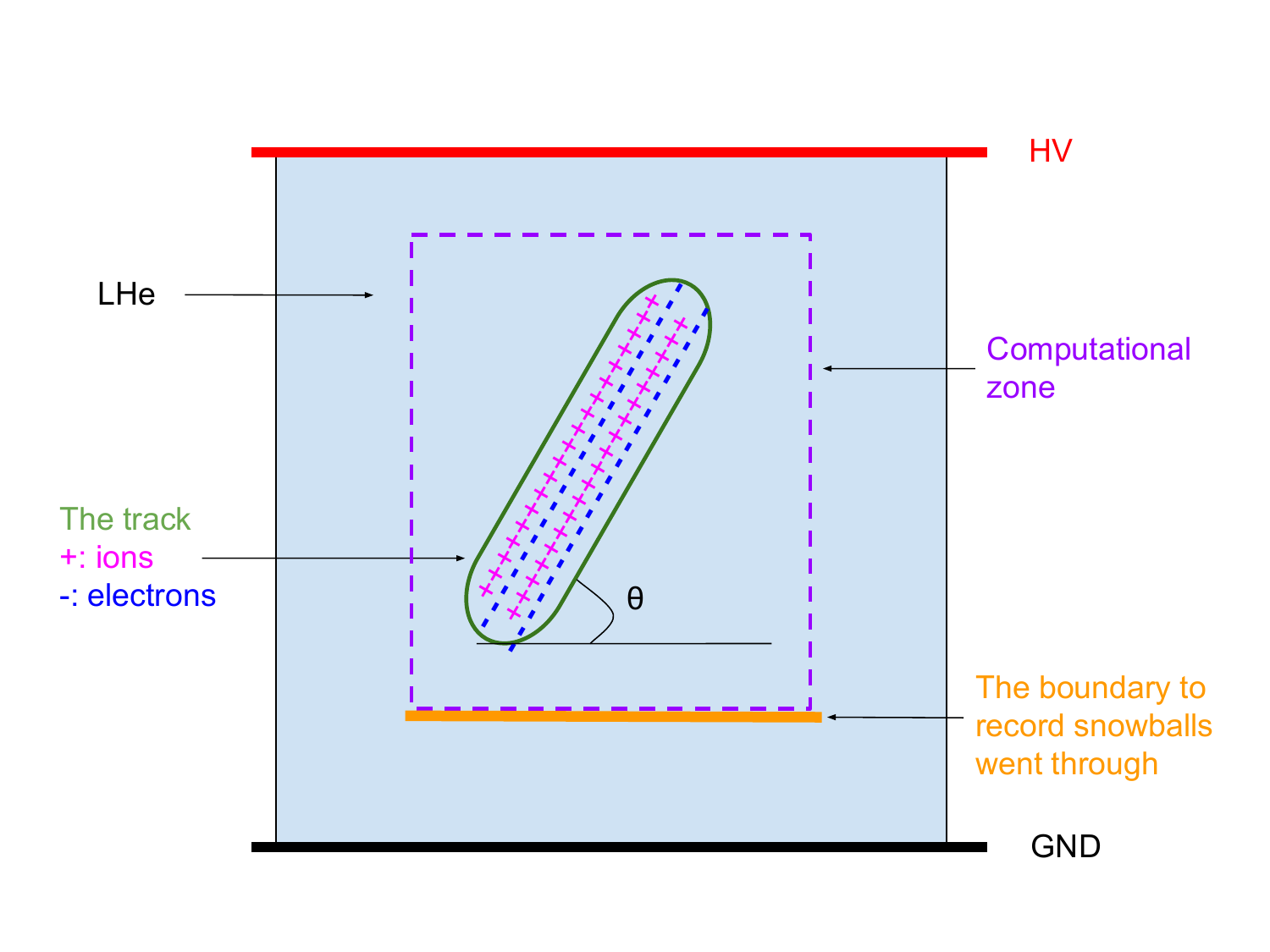}
	\caption{Schematic of the simulation setup: an alpha track in LHe under an external electric field. The computational zone marks the simulation region. The bottom boundary is the surface where drifted ions/snowballs are recorded.}
	\label{fig:electronEscapeSchematic}
\end{figure}

Fig.~\ref{fig:electronEscapeSchematic} shows the simulation setup schematically. An alpha-generated track is oriented at an angle $\theta$ with respect to the electric field. The computational zone, which is smaller than the full electric field region, is used to reduce the computational load. Any electron/bubble or ion/snowball that exits this zone is considered unable to recombine further, given the presence of the external field 10 kV/cm~\footnote{In the simulation, when a snowball reaches the boundary, the minimum distance between it and its nearest bubble is approximately 300 nm. At the midpoint between the two particles, the electric field is about 1.2 kV/cm, which is roughly eight times weaker than the applied external field of 10 kV/cm. As a result, the snowball and bubble will continue to separate under the external field and can no longer recombine.}. The boundary at the bottom of the computational zone is used to record the ions/snowballs that pass through it.

Our simulations were performed in three steps. Step 1: we simulated the 5.3 MeV alpha track under 10 kV/cm in 1.41 K LHe to obtain results consistent with experimental data. After accomplishing Step 1 successfully, we proceeded to Step 2: applying the same simulation setups to keV-scale alpha particles. Step 3: extracting a relationship between the escape ratio and the number density of the ionized particles.

The manuscript is organized as follows. In Section~\ref{sec:implementSimulation}, we introduce the physical process of an energetic alpha particle in LHe. We present the simulation details in Section~\ref{sec:realSimulation}. We report the simulated results in Section~\ref{sec:simulationResults} and discuss the obtained results in Section~\ref{sec:simulationDiscussions}.

\section{Physical processes to be simulated} \label{sec:implementSimulation}

An energetic alpha particle interacts with LHe atoms primarily through two processes: excitation and ionization. The excitation energy and ionization energy of helium are 20.6~eV~\cite{Jesse55} and 24.6~eV~\cite{Smirnov82}, respectively. However, the average energy required to produce one electron–ion pair (the W-value) is approximately 42.3~eV~\cite{Jesse55}; the energy discrepancy of approximately 11.7~eV is converted into the electron's initial kinetic energy and the heat of the liquid.

In general, the ionized electrons possess initial kinetic energies ranging from tens to hundreds of eV. They subsequently thermalize on a picosecond (ps) timescale through both inelastic collisions, which rapidly reduce their kinetic energy, and elastic collisions with surrounding helium atoms. When the electron kinetic energy drops below approximately 1~eV, the electron pushes away surrounding helium atoms due to Pauli exclusion, forming a bubble; the measured timescale for this process is about 4~ps~\cite{RosenblitAndJortner95}. Meanwhile, the positive ions He$^+$ produced by ionization attract nearby helium atoms through polarization forces, forming a snowball structure within a very short time (approximately 1~ps). After the formation of bubbles and snowballs, they drift under the influence of the external electric field and may undergo recombination. The recombination process dominates on a timescale of ns~\cite{McKinsey2003}.

In the simulation, electrons are initialized at T$_0$ with an initial energy of 0.95~eV, corresponding to the state just before bubble formation. Ions are initialized as snowballs at the beginning with no initial energy. The simulation then studies the subsequent drift, diffusion, and recombination of bubbles and snowballs~\footnote{For convenience, we may use ``electrons'' to refer to bubbles and ``ions'' to refer to snowballs in the remainder of the manuscript.} in LHe under a uniform external electric field.

Specifically, the main physical processes included in the simulation are as follows:\\
(a) Electron elastic collisions: electrons undergo elastic scattering with liquid helium atoms, which dissipates their residual kinetic energy. Following an elastic collision, if the electron's energy falls below 1 eV, its status is changed to that of a bubble.\\
(b) Recombination of bubbles and snowballs.\\
(c) Drift and diffusion: under a uniform external electric field (e.g., 10~kV/cm), bubbles and snowballs drift in opposite directions.\\
(d) Escape definition: a bubble or snowball is considered to have escaped when it reaches a defined boundary located far from the initial track position (as shown in Fig.\ref{fig:electronEscapeSchematic}), beyond which it cannot return under the external field. The electron escape ratio is then defined as the number of snowballs that reach this boundary divided by the initial number of ions.

\section{Simulation details}\label{sec:realSimulation}

\subsection{COMSOL introduction}

The simulations in this work were performed using the COMSOL Multiphysics 6.3 platform. COMSOL~\cite{comsoWebsite} is a commercial software package based on the finite element method, which solves partial differential equations to simulate physical processes and is particularly well-suited for handling multiphysics coupling problems. Its core principle involves discretizing the entire geometric domain into a large number of small mesh elements, approximating the physical fields within each element using simple polynomial functions, and then assembling the equations from all elements into a large system of linear algebraic equations for solution. In addition to the base multiphysics module, we employed the Plasma Module~\cite{comsol_plasma63} to perform the simulations. The core physical processes to be implemented are as follows.

(i) Electron Drift Diffusion Interface: Instead of solving the Boltzmann equation, electron transport in the Plasma Module is approximated by fluid equations, which are obtained by multiplying the Boltzmann equation by a weighting function and then integrating over velocity space. The parameters of the fluid equations are the electron number density, the mean electron momentum, and the mean electron energy.

(ii) Heavy Species Transport Interface: The transport of neutral atoms, molecules, and ions (heavy species) is well modeled using mass fractions, drift–diffusion, and convection. The module also handles gas-phase reactions, surface reactions, and electron impact source terms for non-equilibrium and thermal discharges. In our simulation, this interface is primarily used to track the distribution of bubbles and snowballs.

(iii) Electrostatics Interface: This interface is used to compute the electric field, electric displacement field, and potential distributions in dielectrics in the presence of bubbles and snowballs. The total electric field, which consists of both the externally applied field and the field induced by the bubbles and snowballs, serves as the driving force for the drift motion of electrons/bubbles and ions/snowballs.

These physical field interfaces are automatically coupled with one another through the steering of the Multiphysics nodes. For example, the ionization and recombination rates computed by the Heavy Species Transport Interface serve as source terms for the Drift Diffusion Interface, while the electric field computed by the Electrostatics Interface directly drives the migration terms in both the Drift Diffusion and Heavy Species Transport Interfaces. The scripts simulate the complete dynamics of particles along the track, from generation, migration, diffusion, recombination, and drift, to eventual collection, which is essential for the accurate simulation of the electron escape ratio.

\subsection{Parameters for relevant physical processes}

The simulated species include electrons, bubbles (denoted as He$^{-}$), and snowballs (denoted as He$^{+}$). The internal reaction parameters and particle transport parameters considered in the simulation are described below.

\begin{figure}[h]
	\centering
	\includegraphics[width=0.9\textwidth]{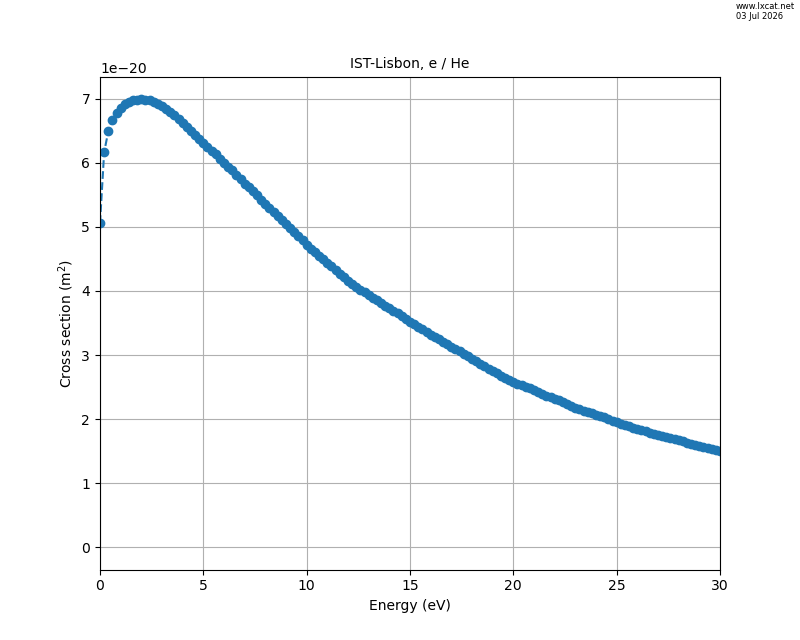}
	\caption{Elastic collision cross section between electrons and LHe as a function of electron energy in the range of 0--30~eV~\cite{lxcat_elastic_he}}
	\label{fig:elastic_cross_section}
\end{figure}

The elastic scattering between electrons and liquid helium atoms is represented by the reaction $\mathrm{e}^{-} + \mathrm{He} \rightarrow \mathrm{e}^{-} + \mathrm{He}$. The cross-section is reproduced in Fig.~\ref{fig:elastic_cross_section} with the original data taken from the LXCat database~\cite{lxcat_elastic_he}.  

\begin{figure}[h]
	\centering
	\includegraphics[width=0.9\textwidth]{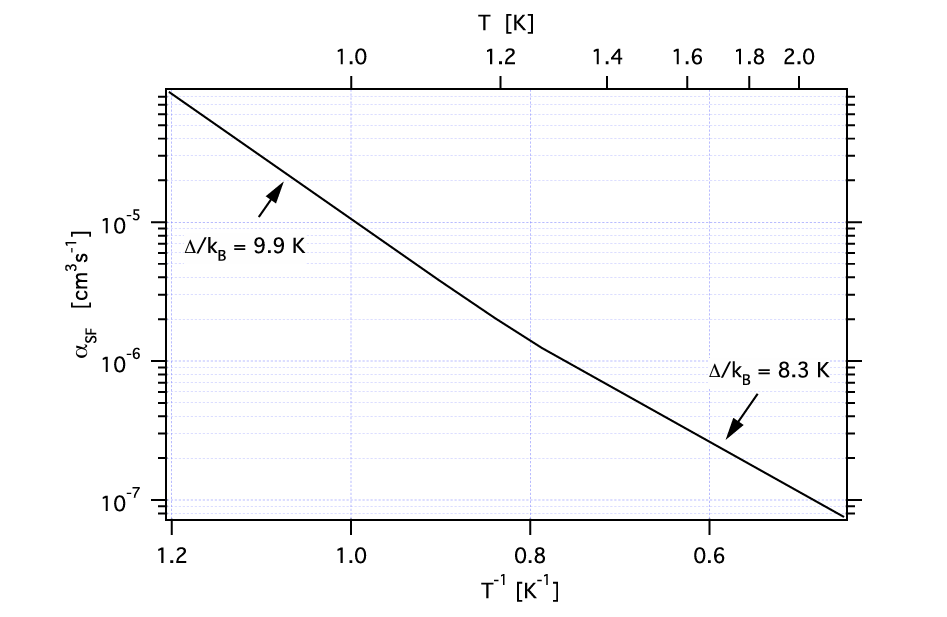}
	\caption{Recombination coefficient $\alpha_{\mathrm{SF}}$ of ions in LHe as a function of inverse temperature $1/T$~\cite{Careri1961}}
	\label{fig:recombination}
\end{figure}

The recombination coefficient of bubbles and snowballs was measured by Careri and Gaeta~\cite{Careri1961} and is reproduced in Fig.~\ref{fig:recombination}. In COMSOL, however, the input parameter for recombination is the so-called ``forward rate constant'' $k^f$, which has the unit of $\mathrm{m^3\,s^{-1}\,mol^{-1}}$ and  is related to $\alpha_{\mathrm{SF}}$ as shown in Eq.(\ref{eq:recombination}). 
\begin{equation}
	\alpha_{\mathrm{SF}} = \frac{k^f}{N_A},
	\label{eq:recombination}
\end{equation}
where $N_A$ is the Avogadro constant.

\begin{figure}[h]
	\centering
	\includegraphics[width=0.9\textwidth]{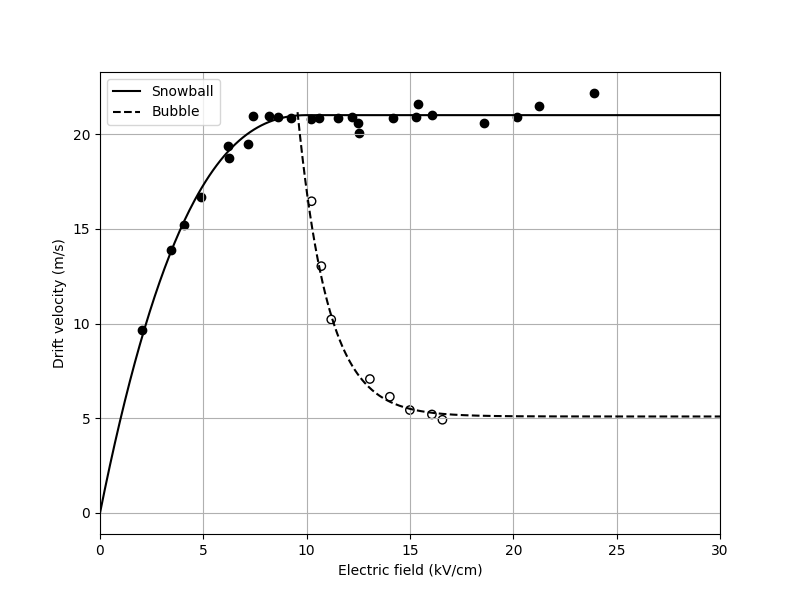}
	\caption{Drift velocity of bubbles and snowballs in LHe at 1.41~K as a function of the electric field~\cite{Bruschi1968}}\label{fig:velocityin1p41K}
\end{figure}

As mentioned above, electrons in LHe thermalize and form bubbles once their energy drops below approximately 1 eV, while ions similarly form snowballs. Given that the mobilities of both bubbles and snowballs depend on the temperature of LHe and the applied external field, we chose to use measured velocities to extrapolate the mobilities, which are the required input parameters in COMSOL. Specifically, the mobilities used in our simulation are based on the measured velocities of bubbles and snowballs at 1.41 K under 10 kV/cm, as shown in Fig.~\ref{fig:velocityin1p41K}, which were reproduced from reference~\cite{Bruschi1968}.

\subsection{Parameters for the $\alpha$ tracks}

A 5.3 MeV $\alpha$ particle produces a cylindrical track in liquid helium with a radius of approximately 60 nm and a length of 270 $\mu$m~\cite{Jaffe1913, Kramers52, ito_effect_2012}. The distance between neighboring ions is around 2 nm~\footnote{This 2 nm spacing can also be estimated from the total number of ions produced by the alpha particle, 1.25E5 = 5.3 MeV / 42.3 eV, where 42.3 eV is the average energy required to produce an electron–ion pair in LHe, and the track length of 270 $\mu$m~\cite{AstarWebsite}; 270 $\mu$m / 1.25E5 $\approxeq$ 2.16 nm.}~\cite{ito_effect_2012}. Thus, the ion distribution along the track forms a tube with a very thin shell of approximately 2 nm in thickness~\footnote{Please refer to Fig.~\ref{fig5p3MeVTrack}}. However, the electrons from the track forms a cylindrical distribution, because the ions' radius is 60 nm and the ionized electrons' range is 100 nm~\cite{Benderskii1999}~\footnote{Please refer to Fig.~\ref{fig5p3MeVTrackElectron}}.  

To address the number density of the two tubes, we implement a Gaussian distribution in cylindrical coordinates, as shown in Eq.(\ref{eq:initial_density}). The electron and ion number densities can be expressed as follows: radially, they follow a Gaussian-shaped distribution centered at the track radius $r_0$ with a standard deviation $\sigma$; axially, they are uniformly distributed over the track length $H$. The mathematical expression is:

\begin{equation}
	n(r,z) = n_0 \exp\left[-\frac{(r-r_0)^2}{2\sigma^2}\right], \quad -H/2 \le z \le H/2,
	\label{eq:initial_density}
\end{equation}
where n (r, z) is the number density of the particles, $n_0$ is the initial charge number density of electrons or ions at the beginning of the simulation, $r_0$ is the center of the track, and $\sigma$ is the standard deviation of the particles's positions. We set $\sigma_e = 35~\mathrm{nm}$ for electrons so that $3\sigma_e$ corresponds to the ionized electron range of 100 nm, and $\sigma_{\mathrm{ion}} = 2~\mathrm{nm}$ for ions, which corresponds to the distance between neighboring snowballs. We tested values of 1 nm and 3 nm for $\sigma_{\mathrm{ion}}$, and the simulation results showed the difference of less than 1\%. We therefore set it to 2 nm. For more information on the parameters, please refer to Table~\ref{tab:comsol_params}.

For keV-scale $\alpha$ tracks, some simulation parameters differ from those used for 5.3 MeV tracks. For instance, the quenching factor (QF)\footnote{The quenching factor quantifies the fraction of recoil energy that can be converted into ionization energy.} is approximately 1 for 5.3 MeV $\alpha$s, while it is only about 0.25–0.6 for keV-scale tracks~\cite{Muraz2016}.

As mentioned above, the radius of 5.3 MeV $\alpha$ tracks in LHe has been determined; however, no similar data exist for keV-scale tracks. Fortunately, Reference~\cite{Khayrat1999} measured the diameters of 0.1–5.3 MeV $\alpha$ particles in CR-39 (allyl diglycol carbonate), from which we can extrapolate the keV track radius in LHe. Unlike our case, the tracks in CR-39 were intentionally enlarged by a 2-hour etching process; consequently, the track diameters in CR-39 are on the scale of $\mu$m, tens of times larger than the diameter of a 5.3 MeV $\alpha$ track in LHe (without etching), which is 120 nm.

The key assumption underlying the extrapolation is that an $\alpha$ track's diameter is proportional to dE/dx, and that this proportionality holds not only within the same medium (CR-39 or LHe) but also across the two media, as shown in Eq.(\ref{eq:diameterOverdEdxSameMed}) and (\ref{eq:diameterOverdEdxDifMed}), respectively.

	\begin{equation}
		\frac  {D_{\mathrm{LHe}}( @Ene\_A ) } 
		{ \left.\mathrm{d}E/\mathrm{d}x\right|_{\mathrm{LHe}}(@Ene\_A) }  =  
		\frac { D_{\mathrm{LHe}}(@Ene\_B) } 
		{\left.\mathrm{d}E/\mathrm{d}x\right|_{\mathrm{LHe}}(@Ene\_B)} 
		\label{eq:diameterOverdEdxSameMed},
	\end{equation}
where ${D_{\mathrm{LHe}}(@Ene\_A ) } $ and  ${ D_{\mathrm{LHe}}(@Ene\_B) }$  are the track diameters in LHe for alpha particles of energies $Ene\_A$ and $Ene\_B$, respectively; and $\left.\mathrm{d}E/\mathrm{d}x\right|_{\mathrm{LHe}}(@Ene\_A)$ and $\left.\mathrm{d}E/\mathrm{d}x\right|_{\mathrm{LHe}}(@Ene\_B)$ are the corresponding dE/dx values in LHe.

	\begin{equation}
		\frac  {D_{\mathrm{LHe}}( @ArbEne ) } 
		{ \left.\mathrm{d}E/\mathrm{d}x\right|_{\mathrm{LHe}}(@ArbEne) }  =  
		\frac{ D_{\mathrm{CR}}(@ArbEne) } 
		{ \left.\mathrm{d}E/\mathrm{d}x\right|_{\mathrm{CR}}(@ArbEne) } 
		\label{eq:diameterOverdEdxDifMed},
	\end{equation}

where $ D_{\mathrm{LHe}}(@ArbEne)$  and  $D_{\mathrm{CR}}(@ArbEne)$  are the track diameters in LHe and CR-39, respectively, for alpha particles of arbitrary energy; and $\left.\mathrm{d}E/\mathrm{d}x\right|_{\mathrm{LHe}}(@ArbEne)$ and $\left.\mathrm{d}E/\mathrm{d}x\right|_{\mathrm{CR}} (@ArbEne)$ are the corresponding stopping powers in the two media.\\

The detailed extrapolation proceeds as follows.

\begin{figure}[h]
	\centering
	\includegraphics[width=0.9\textwidth]{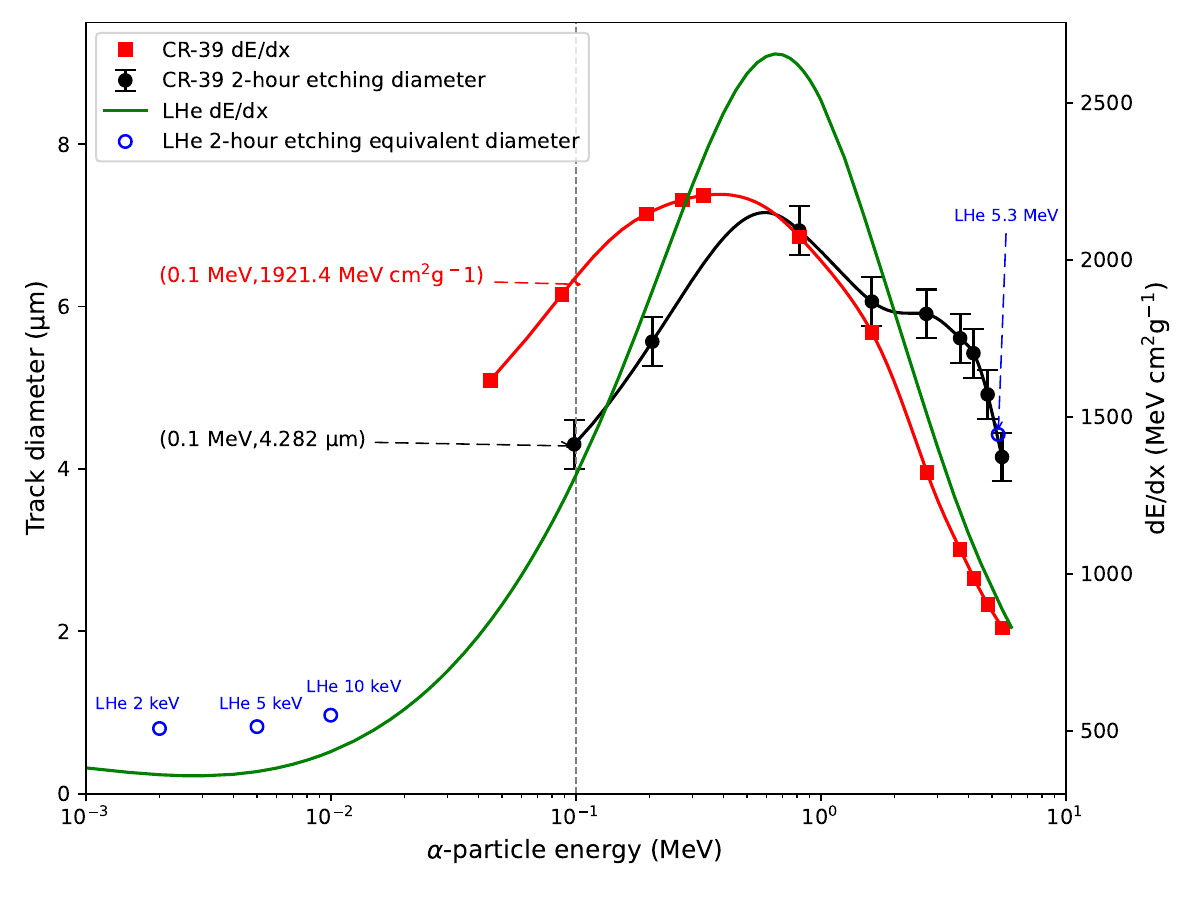}
	\caption{The dE/dx values for CR-39~\cite{Henke1968} and LHe~\cite{AstarWebsite} are shown as the red and green curves, respectively. The black data points represent the track diameters in CR-39 after two hours of etching~\cite{Khayrat1999}. The blue circles denote the extrapolated $\alpha$ track diameters in LHe, assuming the same two-hour etching procedure as in Ref.~\cite{Khayrat1999} was applied. The radii of the unetched $\alpha$ tracks in LHe can be obtained by dividing by the etching scale factor, as summarized in Table~\ref{tab:trackParameters}. For further details, please refer to the main text.}
	\label{fig:track_CR-39}
\end{figure}

(a) Based on the dE/dx of 5.3 MeV $\alpha$ in the two media~\cite{AstarWebsite, Henke1968}, we calculate the hypothetical diameter of the track generated by 5.3 MeV alphas in LHe, assuming the same 2-hour etching procedure was applied as for CR-39 in Ref.~\cite{Khayrat1999}. Using  $D_{\mathrm{CR}}$  (5.3 MeV)= 4.115 $\mu$m,  $\left.\mathrm{d}E/\mathrm{d}x\right|_{\mathrm{LHe}}$(5.3 MeV) = 916.6 MeV $\cdot$ cm$^2$/g and $\left.\mathrm{d}E/\mathrm{d}x\right|_{\mathrm{CR}}$  (5.3 MeV)= 854.5 MeV $\cdot$ cm$^2$/g that can be found in Fig.~\ref{fig:track_CR-39}, we obtain ${D_{\mathrm{LHe}}(\text{5.3 MeV}) } $ = 4.425 $\mu$m.

(b) We then determine the etching scale factor, defined as the ratio between the diameter of the hypothetical etched track and that of the unetched track in LHe for 5.3 MeV $\alpha$s, as shown in Eq.(\ref{eq:etchingScaleFactor}).

	\begin{equation}
		\text{Etching~scale~factor} = \frac{D_{\mathrm{LHe}}(5.3\,\mathrm{MeV})} {r_0(5.3\,\mathrm{MeV})} \approx 73.75
	\label{eq:etchingScaleFactor},
	\end{equation}
where ${D_{\mathrm{LHe}}(5.3\,\mathrm{MeV})}$ was calculated in Step (a) as $4.425~\mu\mathrm{m}$, and ${r_0(5.3\,\mathrm{MeV})}$ is the radius of a 5.3 MeV $\alpha$ track, 60 nm~\cite{Jaffe1913, Kramers52, ito_effect_2012}.
We assume that the etching scale factor also applies to keV-scale tracks in LHe, as will be further discussed in Step (d) below.\\

(c)We obtain the hypothetical diameter of tracks generated by 0.1 MeV $\alpha$s in LHe, assuming a 2-hour etching procedure, using Eq.(\ref{eq:diameterOverdEdxDifMed}). Substituting $D_{\mathrm{CR}}$  (0.1 MeV)= 4.282 $\mu$m,  $\left.\mathrm{d}E/\mathrm{d}x\right|_{\mathrm{LHe}}$(0.1 MeV) = 1313.0 MeV $\cdot$ cm$^2$/g and $\left.\mathrm{d}E/\mathrm{d}x\right|_{\mathrm{CR}}$  (0.1 MeV)= 1921.4 MeV $\cdot$ cm$^2$/g into the equation yields ${D_{\mathrm{LHe}}(\text{0.1 MeV}) } $ = 2.926 $\mu$m.\\

(d) We obtain the hypothetical diameters of tracks generated by 2, 5 and 10 keV $\alpha$s in LHe, assuming a 2-hour etching, based on  ${D_{\mathrm{LHe}}(0.1 \text{MeV}) } $ and Eq.(\ref{eq:diameterOverdEdxSameMed}). Taking 2 keV as an example, substituting $D_{\mathrm{LHe}}$  (0.1 MeV)= 2.926 $\mu$m,  $\left.\mathrm{d}E/\mathrm{d}x\right|_{\mathrm{LHe}}$(0.1 MeV) = 1313.0 MeV $\cdot$ cm$^2$/g and $\left.\mathrm{d}E/\mathrm{d}x\right|_{\mathrm{LHe}}$  (2~keV)= 360.3 MeV $\cdot$ cm$^2$/g into Eq.(\ref{eq:diameterOverdEdxSameMed}) yields ${D_{\mathrm{LHe}}(\text{2~ keV}) } $ = 0.803 $\mu$m. Similarly, the hypothetical diameters for 5 and 10 keV are calculated to be ${D_{\mathrm{LHe}}(\text{5~keV}) } $ = 0.825 $\mu$m and ${D_{\mathrm{LHe}}(\text{10~keV}) } $ = 0.967 $\mu$m, respectively. \\

(e) By combining the enlarged keV-scale track diameters from Step (d) with the etching scale factor obtained in Step (b), the true keV-scale track radii can be determined using Eq.(\ref{eq:2keVRadiusCal}). Taking  the 2 keV $\alpha$ track as an example, substituting $D_{\mathrm{LHe}} (2~\text{keV})$ = 0.803 $\mu$m and an etching scale factor of 73.75 yields a radius of 10.89 nm. The radii for 5 and 10 keV tracks are calculated in the same manner to be 11.19 nm and 13.11 nm, respectively.

	\begin{equation}
		r_0(\text{N~keV}) = \frac{ D_{\mathrm{LHe}} (\text{N~keV}) } {\text{etching scale factor} }
	\label{eq:2keVRadiusCal},
	\end{equation}

The extrapolated radii for 2, 5, and 10 keV tracks are shown in Fig.~\ref{fig:track_CR-39}. In Step (d), we calculated the enlarged keV track diameters using the 0.1 MeV data point rather than the 5.3 MeV one, because the 2, 5, and 10 keV points lie on the same side of the Bragg curve as the 100 keV point, whereas the 5.3 MeV point lies on the opposite side.

All track-related parameters are summarized in Table~\ref{tab:trackParameters}. Regarding the electron-ion pairs in the table,  for keV-scale $\alpha$ tracks, all electron--ion pairs are included; for the $5.3~\mathrm{MeV}$ $\alpha$ track, however, only a $1~\mu\mathrm{m}$ segment of the track, corresponding to 456 electron--ion pairs, was simulated due to the limited computing capacity of a PC. We will discuss this further in Section~\ref{sec:simulationDiscussions}.

\begin{table}[htbp]
	\centering
	\caption{The related parameters for 5.3~MeV and keV-scale $\alpha$ tracks in LHe}
	\label{tab:trackParameters}
	\begin{tabular}{@{}lcccc@{}}
		\toprule
		$\alpha$ track energy & 2 keV & 5 keV & 10 keV & 5.3 MeV  \\
		\midrule
		Quenching factor~\cite{Muraz2016}         & 33\%                               & 53\%                               & 60\%                     & 100\%                       \\
		Projected range (g/cm$^{2}$)   & $3.725\times10^{-6}$                & $1.044\times10^{-5}$               & $2.130\times10^{-5}$                & $3.69\times10^{-3}$      \\
		$\mathrm{d}E/\mathrm{d}x|_{\mathrm{LHe}}$ (MeV$\cdot$cm$^{2}$/g)      & 360.3                               & 370.3                              & 433.9                              & 916.6                \\
		Track length ($\mu$m )      & 0.257                               & 0.720                               & 1.47                               & 270                     \\
		Track radius (nm )       & 10.89                               & 11.19                              & 13.11                              & 60                        \\
		Electron--ion pairs     & 15.35                               & 61.63                              & 139.5                             & $1.25\times10^{5}$          \\
		\bottomrule
	\end{tabular}
\end{table}

Table~\ref{tab:comsol_params} lists all simulation parameters for a 10 keV $\alpha$ track with its axial direction perpendicular to the external electric field.

\newpage

\begin{table}[htbp]
	\centering
	\caption{COMSOL simulation parameters for a 10~keV $\alpha$ track perpendicular to the external electric field}
	\label{tab:comsol_params}
	\small
	\setlength{\tabcolsep}{4pt}
	\begin{tabular}{@{}lllll@{}}
		\toprule
		Parameter & Original value & SI value & Description & Source \\
		\midrule
		$L$                 & $1.2\,[\mu\mathrm{m}]$                    & $1.2\times10^{-6}\,\mathrm{m}$       & Computational zone length &Geometry  \\
		$H_{\text{all}}$    & $1[\mu\mathrm{m}]$                                 & $1\times10^{-6}\,\mathrm{m}$         & Uniform electric field length &Geometry  \\
		$L_{\text{all}}$    & $2[\mu\mathrm{m}]$                                 & $2\times10^{-6}\,\mathrm{m}$         & Computational zone length &Geometry  \\	
		$H$                 & $1.47\,[\mu\mathrm{m}]$                            & $1.47\times10^{-6}\,\mathrm{m}$      & Track length &Table~\ref{tab:trackParameters}  \\
		$r_0$               & $13.11\,[\mathrm{nm}]$                             & $1.311\times10^{-8}\,\mathrm{m}$     & Track radius & Table~\ref{tab:trackParameters} \\
		$E_0$               & $10\,[\mathrm{kV/cm}]$                             & $1\times10^{6}\,\mathrm{V/m}$        & Uniform electric field  &Physics  \\
		$\theta$            & $\frac{\pi}{2}$                                    & $1.5708$                             & Angle between field and track &Geometry  \\
		$\sigma_{\mathrm{ion}}$ & $2\,[\mathrm{nm}]$                             & $2\times10^{-9}\,\mathrm{m}$         &Track's ions distribution  & Ref.~\cite{ito_effect_2012} \\
		$\sigma_e$          & $35\,[\mathrm{nm}]$                                & $3.5\times10^{-8}\,\mathrm{m}$       & Track's electrons distribution& Ref.~\cite{Benderskii1999} \\
		$H_{\text{tail}}$   & $20\,[\mathrm{nm}]$                                & $2\times10^{-8}\,\mathrm{m}$         & COMOSOL inline decay const. & COMSOL \\
		$N_0$               & $1\times10^{16}\,[\mathrm{m}^{-3}]$                 & $1\times10^{16}\,\mathrm{m}^{-3}$    & Background charged density & COMSOL \\
		$N_1$               & $1.4232\times10^{26}\,[\mathrm{m}^{-3}]$            & $1.4232\times10^{26}\,\mathrm{m}^{-3}$&Initial ion density & Eq.(\ref{eq:initial_density}) \\
		$N_2$               & $8.45\times10^{21}\,[\mathrm{m}^{-3}]$              & $8.45\times10^{21}\,\mathrm{m}^{-3}$ &Initial electron density & Eq.(\ref{eq:initial_density}) \\
		$T_0$               & $1.41\,[\mathrm{K}]$                               & $1.41\,\mathrm{K}$                   & Temperature & Ref.~\cite{Bruschi1968} \\
		$\rho$              & $0.145\,[\mathrm{g/cm}^3]$                         & $145\,\mathrm{kg/m}^3$               & Liquid helium density & Ref.~\cite{LHe_1998} \\
		$R$                 & $95\,[\mathrm{nm}]$                                & $9.5\times10^{-8}\,\mathrm{m}$       & Refined mesh region radius & COMSOL \\
		$k^f$               & $8.74\times10^{9}\,[\mathrm{m^3/(s\cdot mol)}]$    & $8.74\times10^{9}\,\mathrm{m^3/(s\cdot mol)}$ & Forward recombination rate & Eq.(\ref{eq:recombination}) \\
		$E_{\text{init}}$   & $0.95\,[\mathrm{eV}]$                             & $1.52\times10^{-19}\,\mathrm{J}$     & Initial mean electron energy &COMSOL \\
		\bottomrule
	\end{tabular}
	\begin{tablenotes}[para]
	\end{tablenotes}

\end{table}

\section{Simulation results}\label{sec:simulationResults}

\begin{figure}[h]
    \centering
    \includegraphics[width=0.9\textwidth]{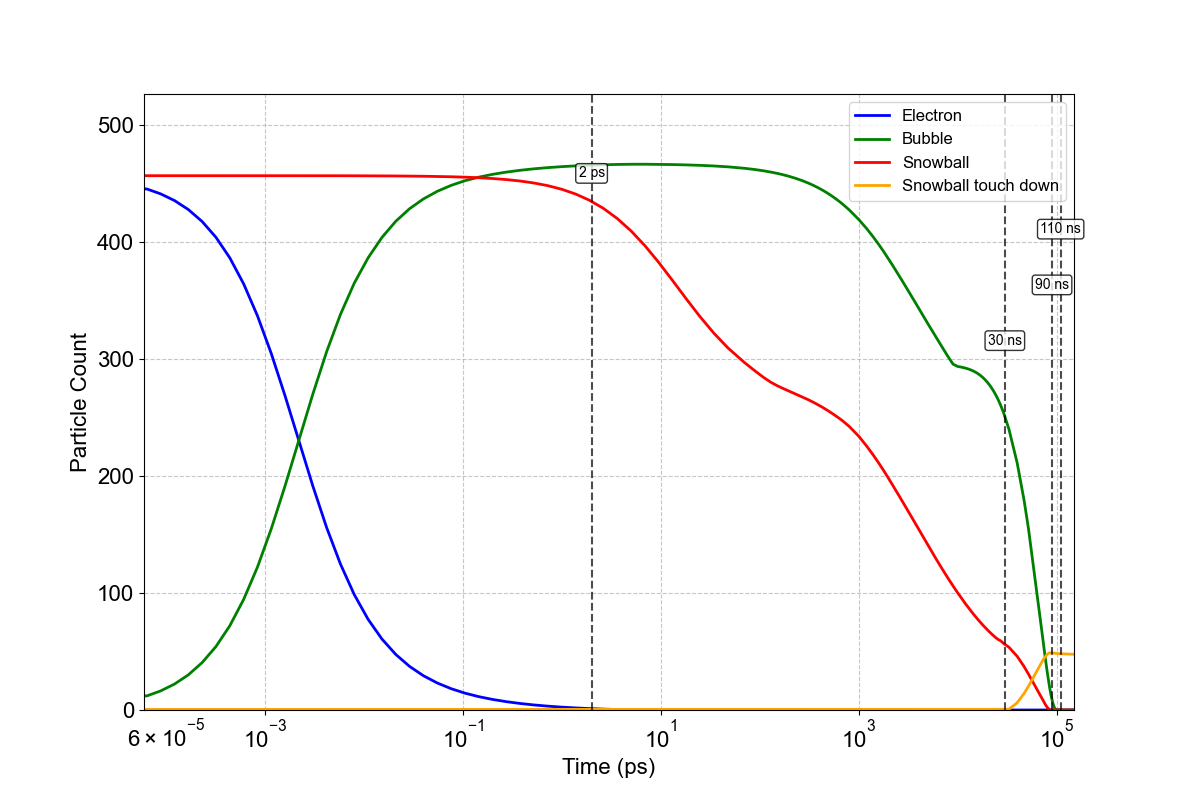}
    \caption{The plot illustrates the simulated particle lifetimes. At T$_0$ + $6 \times 10^{-5}$~ps, 11 electrons have thermalized into bubbles. All electrons are thermalized by $\sim$ 2 ps, when the bubbles count peaks. Bubbles are eliminated by $\sim$ 110 ns via recombination or drift. The 456 snowballs begin recombining with bubbles at  T$_0$ + 0.1 ps and disappear by around 90 ns. Some snowballs reach the boundary at  T$_0$ + 30 ns, and all have passed through within the next 60 ns. See the main text for further details.} 
    \label{fig:ParticlesvsTime}
\end{figure}

COMSOL can export detailed process information for the simulated particles, as shown in Fig.~\ref{fig:ParticlesvsTime}. The plot illustrates the full lifetime of electrons/bubbles and ions/snowballs for a track generated by a 5.3 MeV $\alpha$ particle. Although the simulation starts at T$_0$ = 0 ps, the plot begins at  $6\times10^{-5}$~ps. At this moment, the number of electrons initially at 0.95 eV has decreased from 456 to 445; the difference of 11 corresponds exactly to the number of bubbles generated. The number of bubbles continues to increase as the electron population decreases. All electrons are thermalized into bubbles by approximately 2 ps, which is consistent with the measured time reported in Ref.~\cite{RosenblitAndJortner95}. Around T$_0$ +110 ns, all bubbles have either recombined or drifted out of the simulation zone. The snowballs, which have zero kinetic energy at T$_0$,  begin to recombine with electrons at about 0.1 ps and vanish approximately 90 ns later. Some snowballs reach the boundary around 30 ns, and by 90 ns all snowballs have passed through. Please refer to Fig.~\ref{fig:electronEscapeSchematic} for more information.

The simulated electron escape ratios are shown in Table~\ref{tab:escapeRatio}. As can be seen from the table, for each of the four energies, simulations were performed at four different angles. The mean escape ratio is the final value we adopt. For the 5.3 MeV track, the measured escape ratio at 1.3 K is 9.9\%~\cite{gerritsen1948}, which is consistent with our simulated results at 1.41 K. According to Ref.~\cite{williams1957}, the ratio is temperature independent between 1.4 K and 4.2 K under an external field of 10 kV/cm and below. Therefore, the measured ratio at 1.3 K should be very close to that at 1.41 K, if not identical.

\begin{table}[htbp]
    \centering
    \small
    \caption{The simulated electron escape ratios for 2, 5, and 10 keV as well as 5.3 MeV $\alpha$ tracks in LHe}
    \label{tab:escapeRatio}
    \begin{tabular}{ccccc c}
        \hline
        $\alpha$ track energy & $0^\circ$ & $30^\circ$ & $60^\circ$ & $90^\circ$ 	&mean\\
        \hline
        2 keV   & 22.81\% & 29.42\% & 24.03\% & 23.34\% & 24.90$\pm$3.05\% \\
        5 keV   & 12.22\% & 17.45\% & 17.52\% & 17.25\% & 16.11$\pm$2.60\% \\
        10 keV  & 16.37\% & 20.09\% & 17.60\% & 17.38\% & 17.86$\pm$1.58\% \\
        5.3 MeV & 8.58\%  & 10.81\% & 10.92\% & 10.32\% & 10.16$\pm$1.08\% \\
        \hline
    \end{tabular}
\end{table}

\begin{figure}[h]
    \centering
    \includegraphics[width=0.9\textwidth]{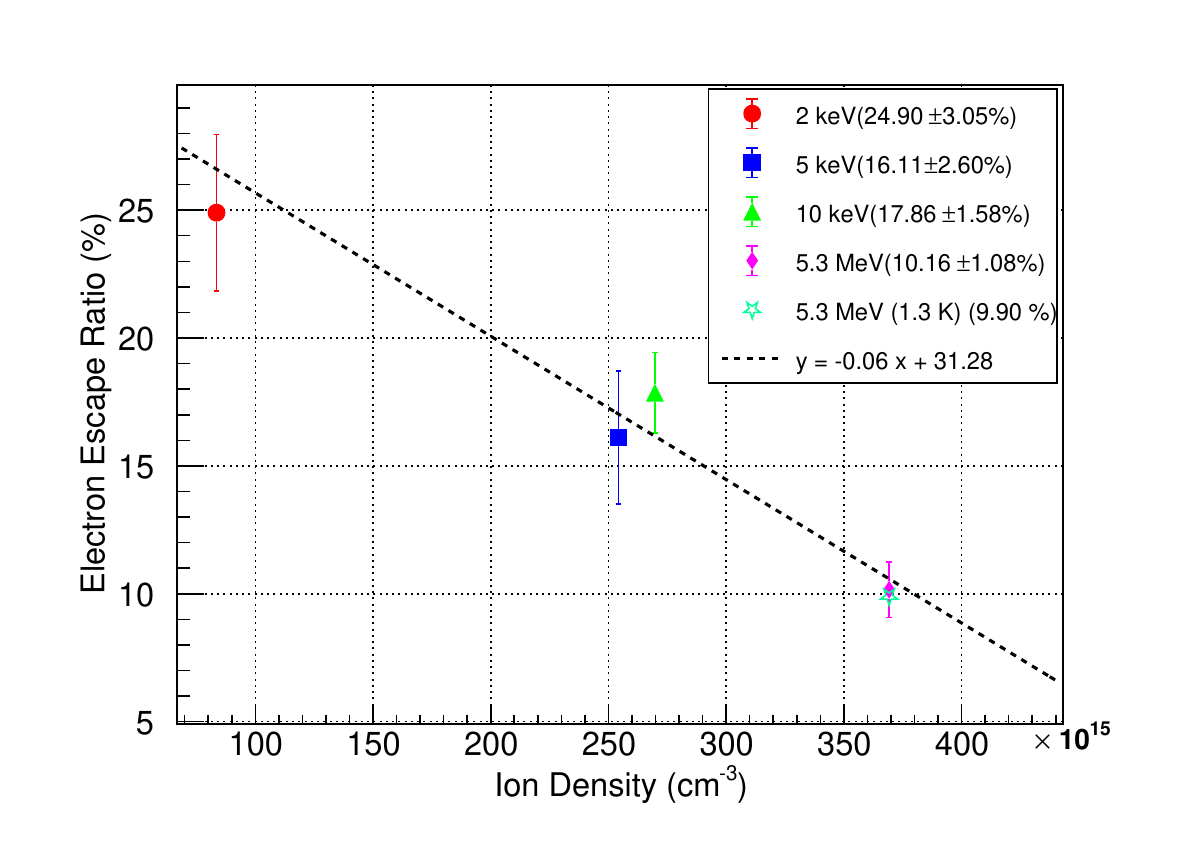}
    \caption{Electron escape ratio versus ion density at T$_0$ for $\alpha$ particles at various incident energies. The cyan open star indicates the experimental value measured at 1.3 K.~\cite{gerritsen1948}.}
    \label{fig:EscapeRateforIonsDensity}
\end{figure}

\begin{figure}	
	\centering
	\begin{subfigure}[t]{2.5in}
		\centering
		\includegraphics[scale=0.15]{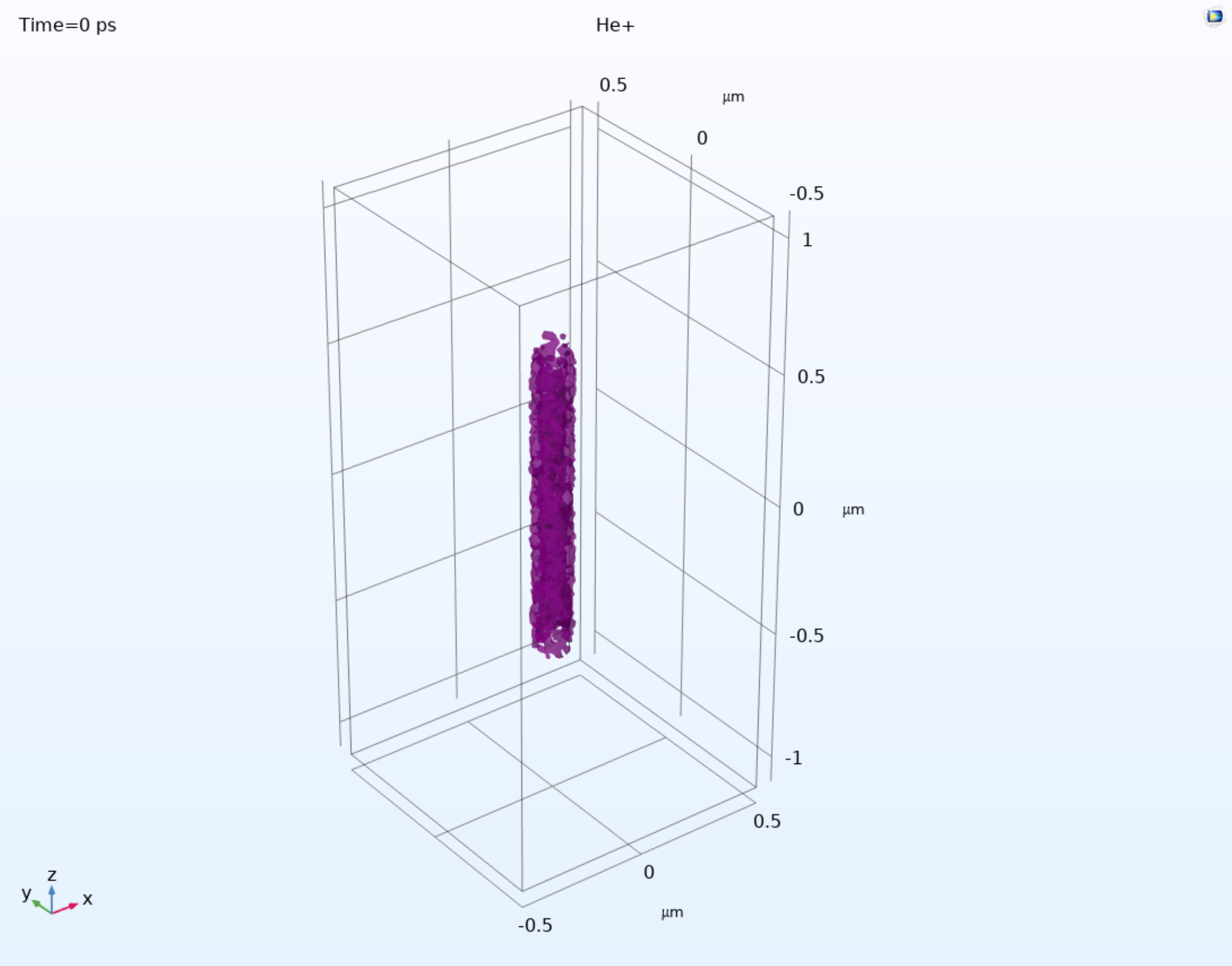}
		\caption{Fig.~\ref{fig5p3MeVTrack.a}. Side view of the 5.3 MeV track's ions.} \label{fig5p3MeVTrack.a}	
	\end{subfigure}
	\hfill
	\begin{subfigure}[t]{2.5in}
		\centering
		\includegraphics[scale=0.15]{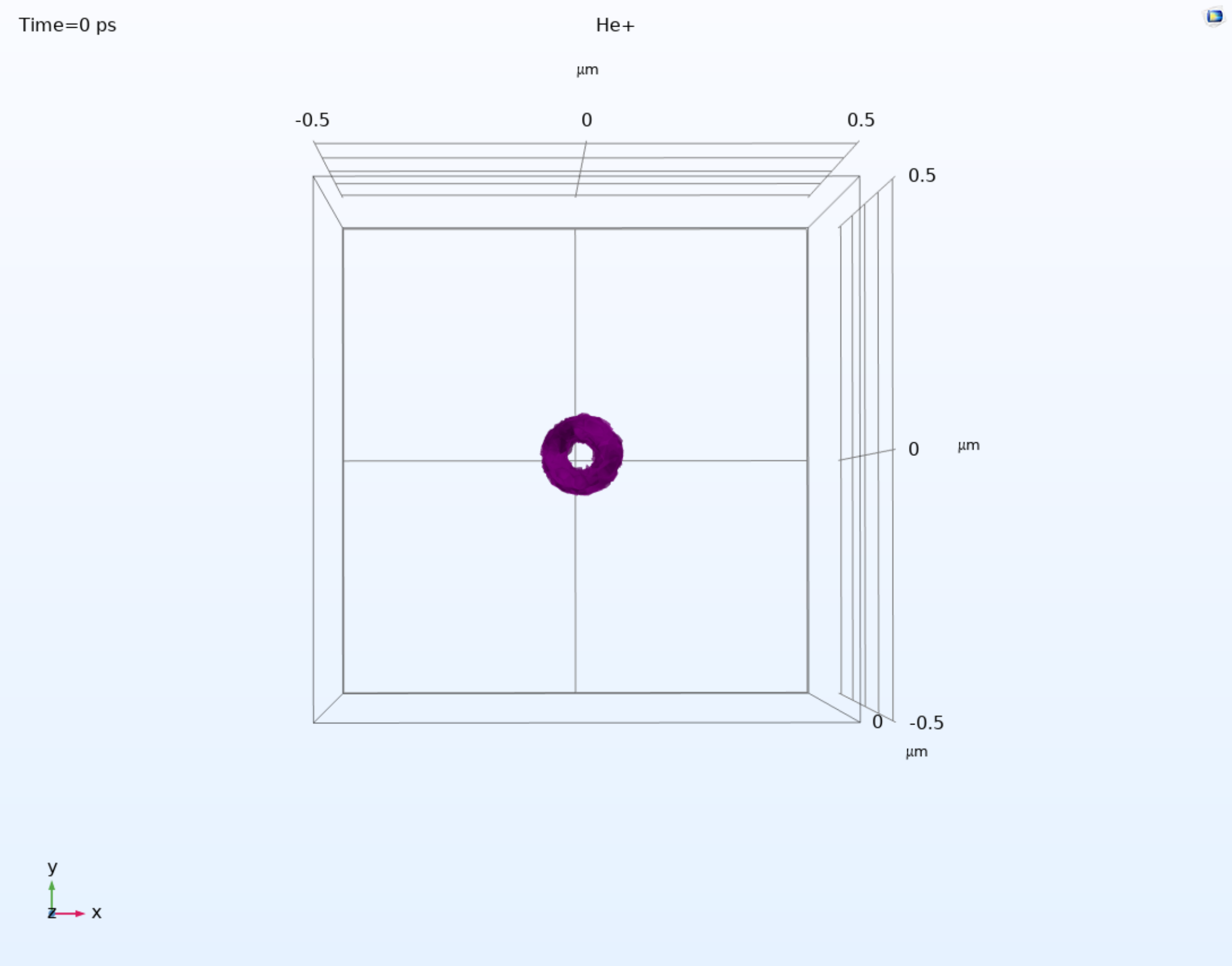}
		\caption{Fig.~\ref{fig5p3MeVTrack.b}. Top view of the 5.3 MeV track's ions.} \label{fig5p3MeVTrack.b}
	\end{subfigure}
	\caption{The spatial distribution of 99\% of the ions for the 5.3 MeV $\alpha$ track at T$_0$.}\label{fig5p3MeVTrack}
\end{figure}

We further investigated the origin of the escape ratio behavior shown in Table~\ref{tab:escapeRatio}. Our analysis reveals a linear relationship between the escape ratio and the ion density of the track, as shown in Fig.~\ref{fig:EscapeRateforIonsDensity}. Here, the ion density is defined as the number of ions corresponding to 99\% of the track's total ion population, divided by the volume containing those ions. Both the ion counts and the corresponding volumes were exported from COMSOL. The volume selection was performed manually as follows: we first set a charge density slightly above the background level in COMSOL and checked how many ions were included. If the fraction dropped below 99\%  (from 100\%), we stopped; otherwise, we gradually increased the charge density until the 99\% threshold was reached. This way, we obtained the volume corresponding to 99\% of the ions, as shown in Fig.~\ref{fig5p3MeVTrack.a} and.~\ref{fig5p3MeVTrack.b}.

\captionsetup[subfigure]{labelformat=empty}

\begin{figure}[h]
    \centering
    \includegraphics[width=0.9\textwidth]{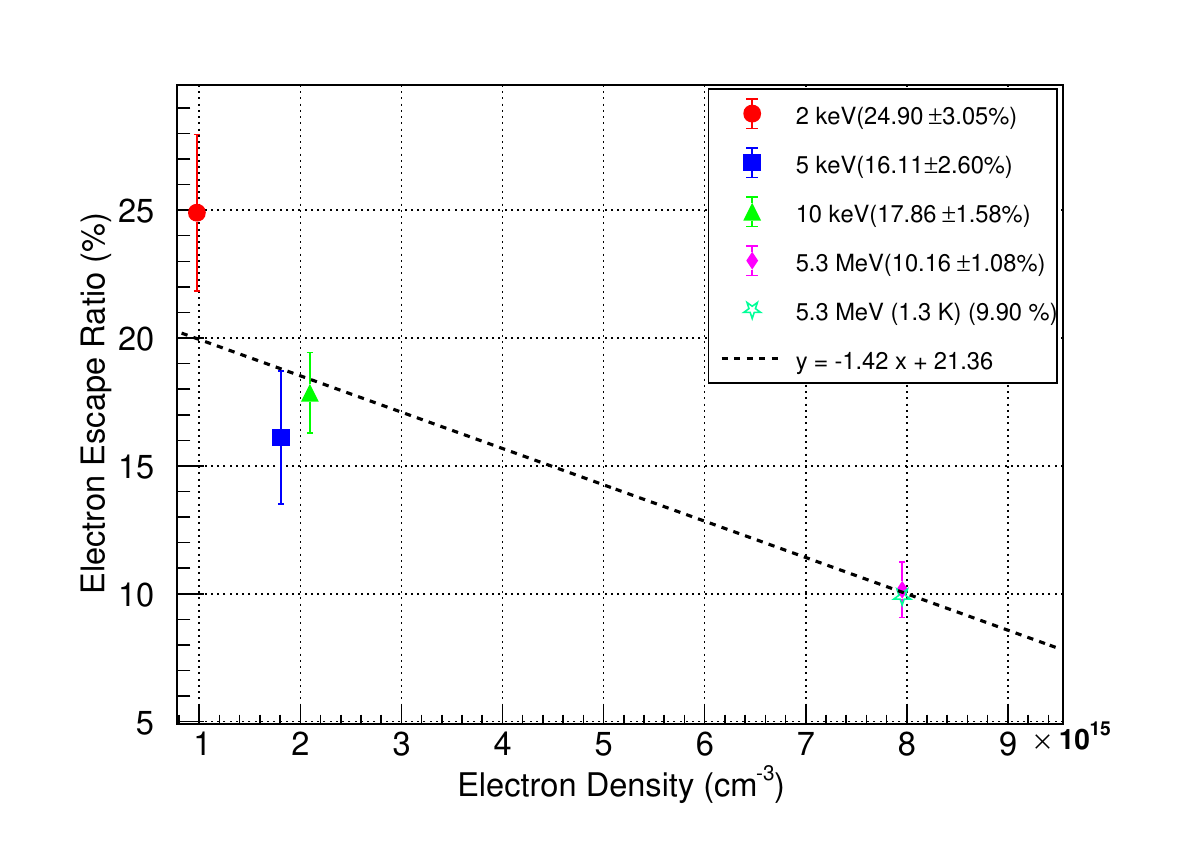}
    \caption{Electron escape ratio versus electrons density at T$_0$ for $\alpha$ particles at various incident energies. The cyan open star indicates the experimental value measured at 1.3 K~\cite{gerritsen1948}.}
    \label{fig:EscapeRateforElectronsDensity}
\end{figure}

\begin{figure}	
	\centering
	\begin{subfigure}[t]{2.5in}
		\centering
		\includegraphics[scale=0.15]{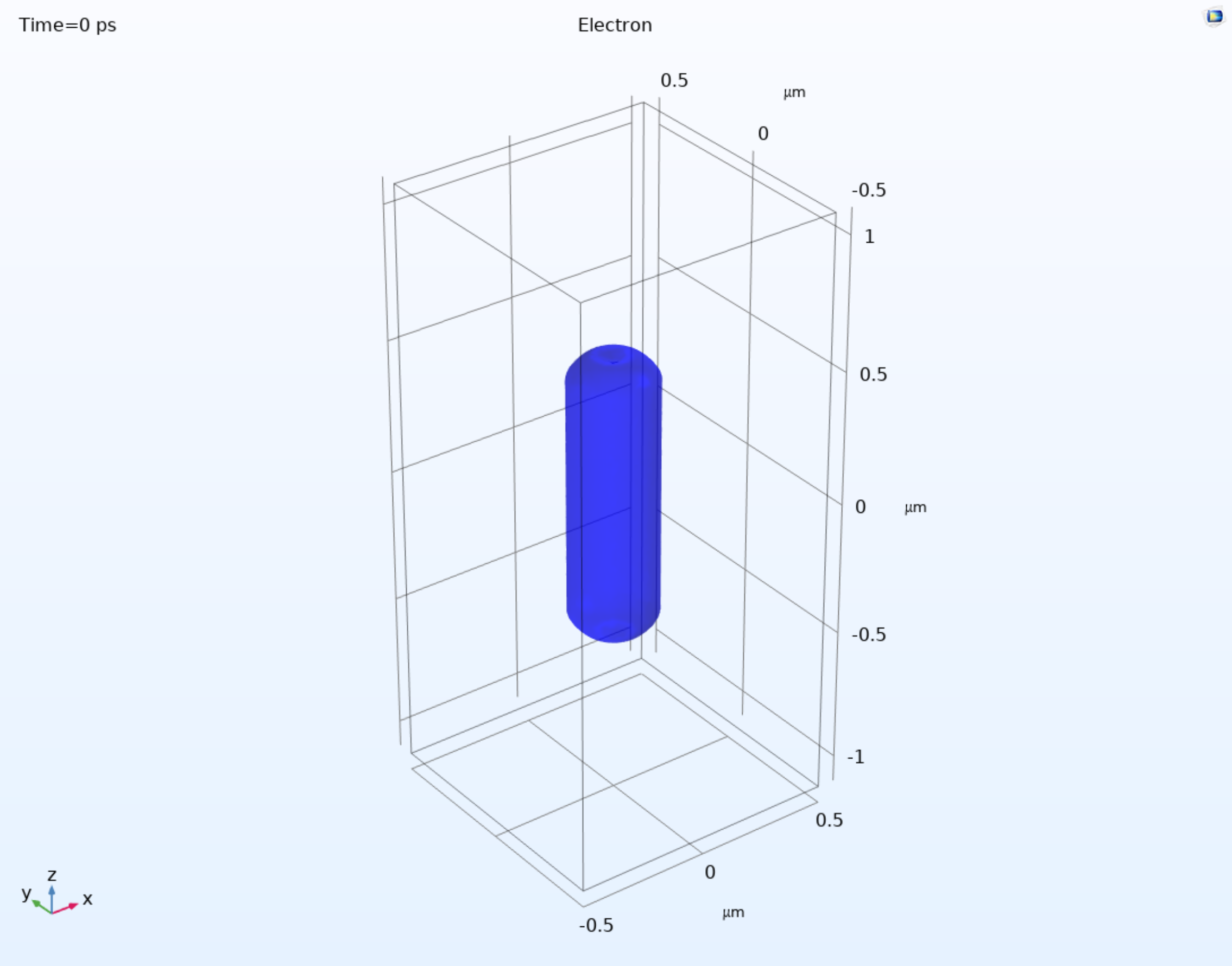}
		\caption{Fig.~\ref{fig5p3MeVTrackElectron.a}. Side view of the 5.3 MeV track's electrons.} \label{fig5p3MeVTrackElectron.a}	
	\end{subfigure}
	\hfill
	\begin{subfigure}[t]{2.5in}
		\centering
		\includegraphics[scale=0.15]{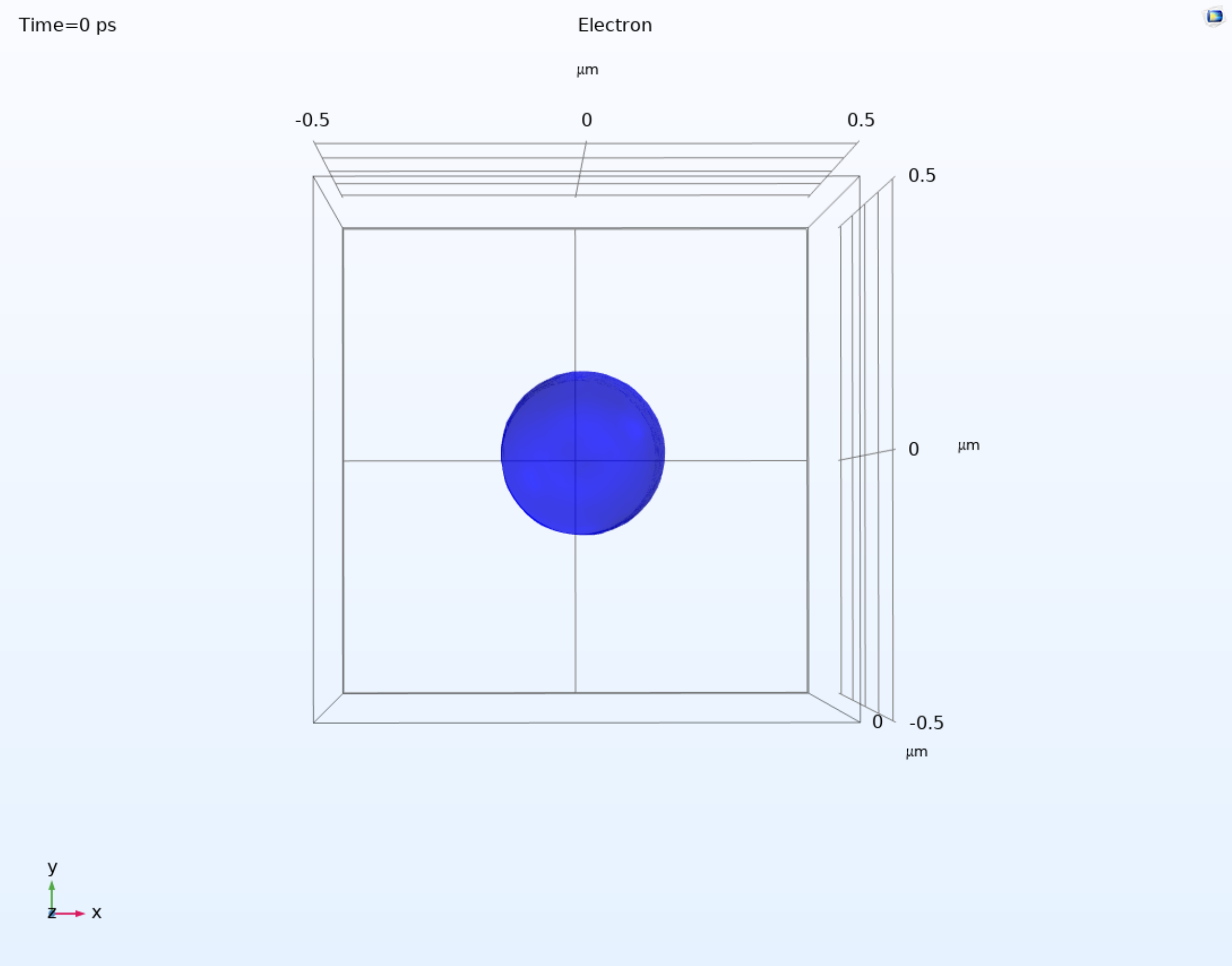}
		\caption{Fig.~\ref{fig5p3MeVTrackElectron.b}. Top view of the 5.3 MeV track's electrons.} \label{fig5p3MeVTrackElectron.b}
	\end{subfigure}
	\caption{The spatial distribution of 99\% of the electrons for the 5.3 MeV $\alpha$ track at T$_0$.}\label{fig5p3MeVTrackElectron}
\end{figure}

We performed the same analysis for electrons but did not observe a similar linear relationship as seen for ions, as shown in Fig.~\ref{fig:EscapeRateforElectronsDensity}. Our preliminary interpretation is that the ions have a charge density approximately two orders of magnitude higher than that of the electrons; therefore, they play a more decisive role in recombination under a 10 kV/cm external field. The spatial distribution of 99\% of the electrons is shown in Figs.~\ref{fig5p3MeVTrackElectron.a} and~\ref{fig5p3MeVTrackElectron.b}.

\captionsetup[subfigure]{labelformat=empty}

\section{Discussion}\label{sec:simulationDiscussions}

Although our simulated electron escape ratio for 5.3~MeV $\alpha$ generated tracks is in good agreement with experimental data~\cite{gerritsen1948}, only approximately 0.4\% (= 456 / 125{,}000) of the electron-ion pairs were simulated. The consistence may be justified by the cylindrical symmetry of the track: even a 0.4\% segment may be representative of the entire track. We simulated only such a small fraction due to two reasons: (i) the simulation with 456 electron–ion pairs already requires about 100 hours on a PC, making a full-track simulation impractical; and (ii) we currently do not have access to a high-performance computing (HPC) cluster on which both the Multiphysics and Plasma modules are installed. We have been working toward obtaining access to such an HPC system for several months and hope to have it available in the coming months.

Owing to the same limitation in computing resources, we are currently unable to extend our simulations to a wider range of electric fields, for example, from 5 kV/cm to 100 kV/cm, as explored in Ref.~\cite{gerritsen1948}. We would also like to perform simulations at more than the current four angles to investigate any possible correlation between the incident angle and the escape ratio. In addition, we plan to simulate electron recoil (ER) events induced by incident electrons once HPC access becomes available. Experimental validation could be carried out using a facility similar to the one we proposed for nuclear recoil (NR) and ER calibration~\cite{Gu2024}.

\section{Conclusion}

This work simulated the electron escape ratio for tracks generated by 2, 5, and 10 keV as well as 5.3 MeV $\alpha$ particles in liquid helium at 1.41 K under a 10 kV/cm electric field. The simulated escape ratio for the 5.3 MeV track is well consistent with experimental data, and the ratios for the keV-scale tracks are approximately twice that of the 5.3 MeV track. Furthermore, we observed that the escape ratio is linearly proportional to the ion density of the track at T$_0$, while its dependence on electron density is relatively weaker.

\bmhead{Acknowledgements}
This work has been supported both by the National Natural Science Foundation of China (NSFC) under Contract No. 12ED232612001001 and the "Continuous-Support Basic Scientific Research Project" in China. The authors wish to thank Wenping Liu and other library staff at the China Institute of Atomic Energy for their assistance with literature retrieval.

\bmhead{Data Availability Statement}
The datasets generated during and/or analysed during the current study are not publicly available due to a preliminary research stage but are available from the corresponding author on reasonable request.

\bibliography{sn-bibliography}

\end{document}